# Scalability and Dimensioning of Network-Capacity Measurement System using Reflecting Servers

Svante Ekelin
Humlegårdens Ekolager
Formerly with Ericsson Research, Sweden

Andreas Johnsson and Christofer Flinta
Research Area Cloud Technologies
Ericsson Research, Sweden

*Abstract*—In a class of methods for measurement of available path capacity and other capacity-related metrics in a network, trains of probe packets are transmitted from a sender to a receiver across a network path, and the sequences of time stamps at sending and reception are analyzed. In large-scale implementations there may potentially be interference between the probe-packet trains corresponding to several concurrent measurement sessions, due to congestion in the network and common measurement end points.

This paper outlines principles for large-scale deployments of network capacity measurement methods using standardized network functionality. Further, the paper provides an in-depth study of dimensioning and scalability challenges related to the measurement end-points of such systems.

The main result is a framework for dimensioning of large-scale network capacity measurement systems based on TWAMP. The framework is based on a method for explicit calculation of queue-length and waiting-time distributions, where results from M/G/1 queuing theory are combined with Monte Carlo integration.

## I. INTRODUCTION

Active measurements have been used for a long time in the evaluation of network performance, where ICMP [15] has been the most commonly used protocol for IP networks.

One branch of research relates to algorithms and tools for actively measuring parameters such as IP-layer available path capacity [12] and tight section capacity [2].

In this study we outline an architecture for large-scale deployments of network capacity measurement tools using standardized networking technology. The paper specifically pinpoints and analyses the potential scalability challenges related to the measurement path end-points in such systems. There is a risk for unwanted interference between the probe trains corresponding to concurrent measurement sessions, due to congestion in common measurement path end-point queues and outgoing interfaces.

The main result of this study is a framework for dimensioning of standardized distributed network capacity measurement systems based on IETF TWAMP [6] and IETF TWAMP Value-Added Octets [7], by calculating queue-length and waiting-time distributions.

The paper is organized as follows: Section II discusses related work. Section III outlines the functionality of a standardized large-scale network capacity measurement system. In Section IV the scalability challenges related to such a system are discussed. Section V describes a model of the system. In Section VI, a formalism is presented for the analysis of the model. Section VII presents results from a MATLAB implementation of the formalism. Section VIII discusses the method's applicability to dimensioning. The paper is concluded in Section IX.

## II. RELATED WORK

### A. Active Measurements

Active measurements (aka active probing) have long been widely used for determining performance parameters of packet-switched networks. The basic concept is to transmit probe packets from a sender towards a receiver. Each probe packet is time stamped on both sides of the network path. For IP networks this functionality has long been deployed using IETF ICMP [15] in tools such as ping and traceroute. For Ethernet and MPLS networks the functionality is most often based on ITU-T Y.1731 [16]. These protocols are capable of measuring metrics such as round-trip time, jitter and loss.

An extension of active measurements is network tomography where individual link characteristics are inferred from path measurements. These techniques generally using either algebraic or statistical methods and are well described in the seminal work by Vardi [18], and more recent developments in [17]. In [19][20] an online version of network tomography is proposed and evaluated. The method cannot infer individual link statistics but is capable of localizing where a performance degradation occurred.

### B. Network Capacity Measurement Methods

Typically, the available capacity is measured by injecting bursts of probe traffic from a network node to another network node. When the bursts have a traffic rate large enough to create a transient congestion in parts of the network path, a lower traffic rate of the probe packets at the receiving node is observed. The sending and the received rate are then used as input to an algorithm for estimating the capacity metrics.

Most methods use a sender that sends a sequence of packet trains to a receiver. A packet train is a group of packets with a specified sending time interval between each packet. The scheme of time intervals of a train is method dependent. Some methods use different time intervals between each packet in the train (e.g. using increasing time intervals [3][25]). Other methods use the same time interval between each packet in the train, but vary that time interval for different trains [1][4][5]. In this latter type of methods the sending time for each train will therefore vary, typically between lower and upper bounds. The sending time interval corresponds to a probe rate for each train and is typically randomized from a known probability distribution. Thus, the probe rate varies from one train to another, but always between a minimum and a maximum value, which are specific to the measurement session.

More recent advancements in method design are available in for example [26]. For a more complete picture of methods in the area we refer to a survey paper by by Chaudhari and Biradar [24]. Further, the authors of [13] study the challenges of applying available capacity measurement techniques in residential and access networks with low-capacity devices.

*C. Network capacity measurements using a reflecting server*

IETF has standardized several active measurement protocols. One recent and popular protocol is the Two-Way Active Measurement Protocol (TWAMP) [6][11]. TWAMP can be used for measuring parameters such as one-way delay (in both directions), jitter and packet loss. The basic operation of TWAMP is to let a sender inject test packets towards a reflector. When the reflector receives a test packet it is transmitted back to the sender as soon as possible. Each test packet is time stamped upon departure and arrival; both at the sender and reflector host. The metrics can then be calculated from the time stamps and sequence numbers in the packets.

The original TWAMP can be used for estimating capacity metrics, such as available path capacity on the forward path by sending trains of TWAMP packets with different probe rates from the sender to the reflector. However, it is not possible to estimate capacity in the reverse path using the same trains, since such measurements require the traffic rate of the train in the reverse direction typically to be chosen independently from that of the forward direction. Reflection of packets as soon as possible does not allow the necessary mechanism of setting the inter-packet separation on the reverse path. This is resolved in IETF RFC 6802 on TWAMP Value Added-Octets [7] which introduces a buffering feature in the TWAMP reflector. That is, the reflector receives and stores all packets in a train before sending them back to the sender as a new train, with a reverse rate which can be chosen independently of the forward rate. This is illustrated in Figure 1 where $\Delta t''$ is typically different from $\Delta t'$ (in an original TWAMP reflector $\Delta t'' = \Delta t'$).

RFC 6802 specifies how to embed a structure of fields which enables the TWAMP reflector to alter the inter-packet separation of packets in a train to be reflected. It also describes how the TWAMP reflector shall interpret the new structure.

It should be noted that RFC 6802 does not put any requirements on the analysis algorithm used for estimating the available path capacity. That is, most methods developed in academia (of which some are mentioned in Section II.B) can use this standard for their measurements.

III. LARGE-SCALE NETWORK CAPACITY MEASUREMENTS USING STANDARDIZED NETWORKING FUNCTIONALITY

We envision that the basic building blocks in large-scale network capacity measurement systems comprise TWAMP senders and RFC 6802 reflectors. The functionality would be strategically placed to allow selected network paths of interest to be measured. One scenario is depicted in Figure 2 where several TWAMP senders perform measurements towards the same reflector. This could correspond to a scenario where TWAMP sender functionality is implemented in mobile devices, and the common reflector is put in the mobile backhaul of the operator in order to determine the IP-layer performance of the wireless connections. For other scenarios several TWAMP senders and reflectors could co-exist.

A key challenge related to large-scale network capacity measurements is the fact that the measurement methods themselves consume network resources. That is, the overhead in terms of consumed capacity can become critical if the measured paths partly overlap. In [14] it is also shown that overlapping links of measurement session paths may bias metric estimation due to interference in the network. This is in line with previous research studying the interaction between probe packets and ordinary traffic [22]. Further, in [21] it was shown that if the number of probe senders is limited the impact on TCP traffic is kept low.

An important aspect of this challenge, previously not adequately studied in the literature except for a probabilistic approach in [23], is the case where the path reflector endpoints overlap, i.e. the same reflector node is involved in several concurrent measurement sessions with different senders. This becomes increasingly challenging with the size of the measurement system, specifically the number of TWAMP senders; virtualization also adds to the challenge [27]. An example is the mobile backhaul scenario sketched above. Too many concurrent measurements will turn the reflector node into a bottleneck in itself.

IV. THE NODE SCALABILITY CHALLENGE

The main focus of this paper is to study the scalability from the perspective of a reflecting node. Theoretically, if one could control the exact timing of all probe packets from all senders destined to a certain reflector, one could envisage designing a common scheduler so as to avoid interference in the reflector.

However, even if we would have perfect real-time control over all senders, as the size of the distributed measurement system grows, the scheduling problem becomes increasingly hard and eventually in practice intractable. We therefore study the opposite, and much more scalable, approach where there is no central scheduler in the system. That is, the senders will transmit their packet trains independently of each other towards the reflector. During a given time interval, the reflector may then receive several trains with different desired sending rates for the reverse path.

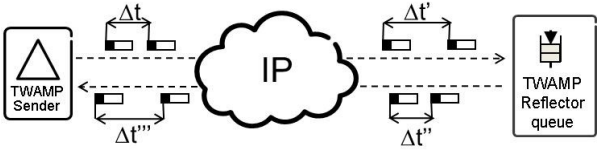

Fig 1. Enhanced TWAMP functionality.

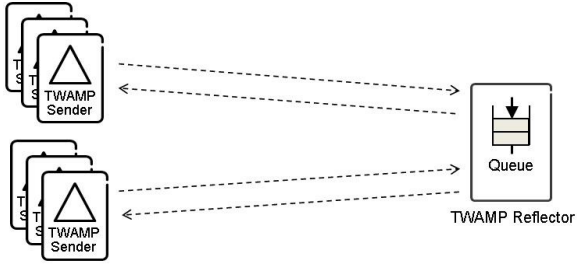

Fig 2. Several distributed TWAMP senders probe the same reflector.

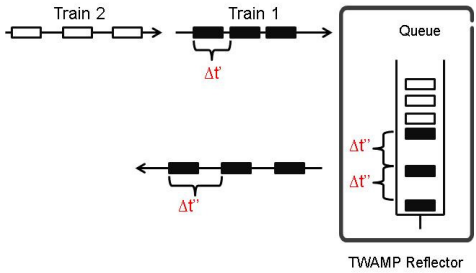

Fig. 3. Illustration of the reflector queue build-up.

To avoid interference between trains sent from the reflector node, each incoming train is put in an out queue and cannot be transmitted until all previous trains in the queue have been sent. *That is, we do not allow for transmission of interleaved trains. Each train blocks the queue handler for the duration of the sending of the packets in the train, including the "gaps" between the packets in the train.* This is due to the time-critical nature of the probe-packet sending times and to the fact that the train rate is randomized, thus making it practically impossible to avoid interference if interleaving would be allowed. This is illustrated in Figure 3. Two trains are arriving at the reflector. Both trains are assumed to be queued before transmission. This results in transmission delay of train 2 due to the sending process of train 1.

The potentially critical condition is the build-up of several trains in the out queue, forcing subsequent trains to experience prolonged waiting times before sending, and possibly being dropped if the buffer depth is not large enough to accommodate them. This is expected to happen more frequently as the total measurement load increases, e.g. when increasing the number of sessions or when increasing the intensity of the sessions.

In order to have a scalable measurement system there must be constraints on the sessions. The main task of the rest of this paper is to dimension the proposed system so that the measurement sessions do not build up queues without bounds on the queueing time at the end points. The critical parameters are: the total number of concurrent sessions, the average number of trains per unit time sent from the TWAMP sender in each session and the sending time durations for each train sent back from the TWAMP reflector in each session. The two first parameters determine the arrival rate to the queue, while the last parameter determines the departure rate from the queue.

In this study we compute the cumulative distribution functions for the out queue length and the waiting time for a train in the queue, for different values of the critical parameters. This will enable us to obtain the quantiles, which can be used for dimensioning guidelines. For example, from requirements on the maximum send time of the trains and requirements on the queue length and waiting time for the 90% quantile, we want to be able to determine the maximum allowed intensity, expressed as the total number of received trains per second. This intensity budget can then be allocated into a number of sessions and session intensity. The maximum waiting time should be low compared to the time scale for each session, i.e. the mean inter-train time separation.

We apply the theory of M/G/1 queues, and use Monte Carlo integration to compute the cumulative distribution function of the queue length and the waiting time. For many cases of practical interest, the queue service time is confined to a sufficiently narrow interval, which makes it feasible to compute the integrals in an explicit expression of the probability density function. We implement the method in MATLAB and present results on dimensioning guidelines.

## V. MODELING THE REFLECTING NODE

We consider $N$ concurrent measurement sessions towards a common RFC 6802 TWAMP reflector. The reflector stores the incoming measurement-packet trains, in the following called probe trains, or simply trains, from the senders in order of arrival in an outgoing queue, before sending reflected trains back in the reverse direction.

Each measurement session is characterized by five independent parameters which have impact on the queuing behavior:

1. Intensity $\lambda$ ( = mean number of probe trains per unit time)
2. Train size $r$ ( = number of probe packets per probe train)
3. Packet size $s$ ( = number of bits per probe packet)
4. Minimum probe rate $u_{min}$ ( = minimum number of bits per unit time for the probe trains sent from the reflector)
5. Maximum probe rate $u_{max}$ ( = maximum number of bits per unit time for the probe trains sent from the reflector)

Each of these parameters can be specific to the measurement session, and when we refer to these specific values we denote the session by a superscript [(n)]. The sessions all contribute to the load of a common queue. Each measurement session $n$ sends probe trains, at a rate $U^{(n)}$. The probe rate is randomly generated for each new train, according to a given probability distribution, and within given bounds

$$u_{min}^{(n)} < U^{(n)} < u_{max}^{(n)} \qquad (1)$$

A uniform distribution is commonly used (e.g. [5]) for capacity measurements, and in the calculations presented in this paper this is what we use for $U^{(n)}$. This method works in principle equally well for any other bounded distribution.

The probe trains are served by the queuing system with a service time, which is the total send time of each train

$$T^{(n)} = \frac{r^{(n)} s^{(n)}}{U^{(n)}} \qquad (2)$$

The cumulative distribution function (cdf) of $T^{(n)}$ can be calculated from the known distribution of $U^{(n)}$:

$$F_{T^{(n)}}(t) = \Pr(T^{(n)} < t) = \Pr(U^{(n)} > r^{(n)} s^{(n)} / t) = \\ 1 - F_{U^{(n)}}(r^{(n)} s^{(n)} / t) \qquad (3)$$

and we then get the pdf of the service time for each arrival process

$$f_{T^{(n)}}(t) = \frac{d}{dt} F_{T^{(n)}}(t) = \frac{d}{dt}(1 - F_{U^{(n)}}(r^{(n)} s^{(n)} / t)) = \\ \frac{r^{(n)} s^{(n)}}{t^2} f_{U^{(n)}}(r^{(n)} s^{(n)} / t) \qquad (4)$$

We also get the bounds for the service time

$$t_{min}^{(n)} < T^{(n)} < t_{max}^{(n)} \qquad (5)$$

where

$$t_{min}^{(n)} = \frac{r^{(n)} s^{(n)}}{u_{max}^{(n)}} \qquad (6)$$

and

$$t_{max}^{(n)} = \frac{r^{(n)} s^{(n)}}{u_{min}^{(n)}} \qquad (7)$$

For the case of uniform probe rate distribution,

$$f_{U^{(n)}}(r^{(n)} s^{(n)} / t) = \frac{1}{u_{max}^{(n)} - u_{min}^{(n)}} \qquad (8)$$

we get a $1/t^2$ behavior for the service time

$$f_{T^{(n)}}(t) = \frac{r^{(n)} s^{(n)}}{u_{max}^{(n)} - u_{min}^{(n)}} \frac{1}{t^2} = \frac{t_{max}^{(n)} t_{min}^{(n)}}{t_{max}^{(n)} - t_{min}^{(n)}} \frac{1}{t^2} \qquad (9)$$

The queuing system under study is modeled as $N$ concurrent independent arrival processes, each with its own intensity $\lambda^{(n)}$ and service time $T^{(n)}$ with distribution $f_{T(n)}(t)$, $(n = 1...N)$.

The pdf of the service time (train sending time $T$) of the whole system can be obtained as a weighted average of the distributions of the $N$ concurrent measurement sessions:

$$f_T(t) = \frac{1}{\lambda} \sum_{n=1}^{N} \lambda^{(n)} f_{T^{(n)}}(t) \qquad (10)$$

where

$$\lambda = \sum_{n=1}^{N} \lambda^{(n)} \qquad (11)$$

is the combined arrival intensity. The proof is given in appendix A. $f_T(t)$ is zero outside the bounded interval $[t_{min}, t_{max}]$, where

$$t_{min} = \min_n t_{min}^{(n)} \qquad (12)$$

$$t_{max} = \max_n t_{max}^{(n)} \qquad (13)$$

If the individual arrival process for each session is Poisson, the combined process is also Poisson. If the individual arrival processes are not Poisson, the combined arrival process may still reasonably well be modeled as a Poisson process, when $N$ is sufficiently large.

The departure intensity $\mu$ of the queue is the inverse of the mean service time

$$\mu = \frac{1}{E(T)} \qquad (14)$$

and the queue traffic intensity $\rho$ is the ratio between the arrival and departure intensities of the queue

$$\rho = \frac{\lambda}{\mu} \qquad (15)$$

VI. SCALABILITY ANALYSIS OF THE REFLECTING NODE

We use the framework of *M/G/1* queue systems to compute the densities of the queue occupancy and the waiting time from known facts about the arrival process and the service time distribution. This is appropriate since the arrival process we are interested in is modeled to be Poisson, (which is memoryless) the service time distribution is general and there is a single server. The service time distributions we consider have the property that the service time is bounded, but apart from that we don't need to make any specific assumptions.

There are some well-known analytical results for the *M/G/1* system, such as the expectation value for the waiting time $W$

$$E(W) = \frac{\lambda E(T^2)}{2(1-\rho)} \qquad (16)$$

This result, which is known as the Pollaczek-Khintchine formula, indicates that the mean waiting time grows beyond all bounds as $\rho$ approaches unity. There are also results for obtaining higher moments of the waiting time, but these are not very suitable for computation of the cdf and quantiles, which is what we are looking for.

One of the well-known results from the M/G/1 theory (see e.g. [8]) is that we have a solution for the steady-state queue occupancy probabilities. The probability for having $i$ elements in the queue is given by the recursive formula

$$\pi_i = \pi_0 k_i + \sum_{j=1}^{i+1} \pi_j k_{i-j+1} \quad (i = 0, 1, 2, ...) \qquad (17)$$

with the arrival probabilities

$$k_i = \int_0^\infty \frac{e^{-\lambda t}(\lambda t)^i}{i!} f_T(t)\, dt \qquad (18)$$

The cdf for the queue length distribution is

$$\Pi_i = \sum_{j=0}^{i} \pi_j \qquad (19)$$

and is derived in Appendix B for completeness. In the following, we use this well-known result and carry it further to arrive at a method for numerical computation of an explicit formula for the waiting time probability density function $f_W(t)$.

For the waiting time $W$, we can express the distribution by conditioning on the queue length. Let $W_i$ be the waiting time given that there are $i$ elements in the queue. We then have

$$f_W(t) = \sum_{i=0}^{\infty} f_{W_i}(t)\, \pi_i \qquad (20)$$

where, for $i \geq 1$,

$$W_i = \sum_{j=1}^{i} T_j \qquad (21)$$

$T_j$ denotes the service time of the $j^{th}$ element in the queue.

For $i = 0$, the case of an empty queue, we have a deterministic distribution: $W_0 = 0$ with probability 1. This means that $W$ has a distribution which is a mixture of a discrete and a continuous distribution. The discrete component only contributes to the value 0.

$T_j$ are independent identically distributed stochastic variables, as given above in (4).

$$f_{T_j}(t) = f_T(t) \qquad (22)$$

The distribution of the sum $W_i$ (with i factors) can then be calculated as a convolution

$$f_{W_i}(t) = f_{T_1} * ... * f_{T_i}(t) = f_T * ... * f_T(t) \qquad (23)$$

For a specific index i we get

$$f_{W_i}(t) = f_T * f_T * ... * f_T(t) = \int_0^\infty ... \int_0^\infty f_T(t - \sum_{j=1}^{i-1} v_j) \prod_{j=1}^{i-1} f_T(v_j)\, dv_1 ... dv_{i-1} \qquad (24)$$

It could seem like a formidable task to compute the waiting time distribution (20) in this formalism, as it involves an infinite series where the terms involve integrals of increasing dimensionality (24). However, for many cases of practical interest this is in fact tractable using a modest amount of computing power. In practice, except for the critical region where $\rho$ is close to 1 (which should be avoided in practical applications since the expectation value of the waiting time tends to infinity as $\rho \rightarrow 1$), the occupancy probabilities $\pi_i$ go rapidly toward zero for increasing $i$. We can then truncate the series and only need to compute the convolution integrals for a finite number of terms in (20), corresponding to $i \leq i_{max}$. For example, $i_{max}$ can correspond to the index $i$ where $f_{wi}(t)\pi_i < \varepsilon$, where $\varepsilon$ is a configurable parameter. That is,

$$f_W(t) = \sum_{i=0}^{\infty} f_{W_i}(t)\pi_i \approx \sum_{i=0}^{i_{max}} f_{W_i}(t)\pi_i \qquad (25)$$

Furthermore, we are interested in a class of applications where $f_T$ is only non-zero on a finite interval *[$t_{min}$, $t_{max}$]*. This means that the generalized integral in (24) reduces to an ordinary integral over a finite region, the hypercube

$$(u_1, \cdots, u_{i-1}) \in [t_{min}, t_{max}]^{i-1}$$

$$f_{W_i}(t) = f_T * f_T * ... * f_T(t) = \int_{t_{min}}^{t_{max}} ... \int_{t_{min}}^{t_{max}} f_T(t - \sum_{j=1}^{i-1} u_j) \prod_{j=1}^{i-1} f_T(u_j)\, du_1 ... du_{i-1} \qquad (26)$$

There is an effective and scalable method for the numerical computation of such multidimensional integrals: the Monte Carlo method [9][10]. In essence, this amounts to randomly generating a sample of points from a distribution uniform over the region of integration, calculating the average value of the integrand from that sample, and multiplying by the size of the region of integration. This approach is simple, effective and scalable, even for large problems the error is well-behaved and the complexity does not explode with the dimensionality.

The computed value of the integral is really an estimate, as this is a statistical method. The larger the sample we use, the higher precision we achieve. The Monte Carlo method also allows an estimate of the variance, thus providing control over the error introduced.

Further, as we are mainly interested in the cdf of the queue waiting time

$$F_W(t) = \int_{-\infty}^{t} f_W(v)\,dv \tag{27}$$

most of the statistical errors from the Monte Carlo estimation of the pdf cancel out in the summation, and we arrive at rather high-precision values for the cumulative distribution function.

## VII. EXPERIMENTAL SETUP AND NUMERICAL RESULTS

We implemented the computation method described above in MATLAB. The input is a scenario consisting of an arbitrary number of measurement sessions, each of them characterized by the five parameters intensity $\lambda$, train size $r$, packet size $s$, minimum probe rate $u_{min}$ and maximum probe rate $u_{max}$.

For a given scenario we successively compute:

- $\lambda$, the aggregated arrival intensity, by (11),
- $t_{min}$ and $t_{max}$, the bounds for the aggregated service time distribution, by (12) and (13)
- $f_T(t)$, the pdf for the service time, by (10)
- $\mu$, the departure intensity, by (14)
- $\rho$, the queue traffic intensity, by (15)
- $k_i$, the arrival probabilities during a service time, by numerical integration of (18) over the discretized interval $[t_{min}, t_{max}]$
- $\pi_i$, the steady-state queue length probabilities, by (17)
- $i_{max}$, the cut-off in queue length used for the waiting-time computation, by the cdf for the queue length distribution from (19) and a configured requirement for probability mass to be included
- $f_{Wi}(t)$, the pdf for the waiting time given that the queue length is $i$, by Monte Carlo integration of the convolution (24)
- $f_W(t)$, the pdf for the waiting time, by a truncated summation of (20)
- $F_W(t)$, the cdf for the waiting time, by (27)

The functions $f_T(t)$, $f_{Wi}(t)$, $f_W(t)$ and $F_W(t)$ are all computed for each point in a discretization of the time interval of interest. This interval is for the service time distribution $[t_{min}, t_{max}]$ and for the waiting time distributions $[t_{min}, i_{max} t_{max}]$.

There are also some computational parameters needed: number of steps in the time interval discretization, required precision in terms of the queue length probability mass included, required precision for the Monte Carlo computation of $f_{Wi}(t)$ in terms of tolerance for deviation from 1 for the integral and initial number of Monte Carlo sample points. These can be set for arbitrary precision in the final result, at the expense of increased need for computation resources.

The MATLAB implementation includes an algorithm which automatically adapts the number of sample points for the Monte Carlo integrations in order to reach the required precision for the integrated $f_{Wi}(t)$ (which ideally should be 1), but we accept $1 \pm \varepsilon$, where $\varepsilon$ is a configuration parameter, typically set to 0.01 in our experiments. In case the requirement is not met, the Monte Carlo integration is re-performed, increasing the number of sample points by a factor of 4. When the requirement is met, the next integral is computed, cutting back the number of sample points by a factor of 2. This way, we ensure that the integral computations have satisfactory precision, while avoiding unnecessarily long execution times for higher-order components.

TABLE I. EXECUTION PARAMETERS AND OUTCOME

| Scenario | Pure 33 | Pure 50 | Pure 66 | Mixed 33 | Mixed 50 | Mixed 66 |
|---|---|---|---|---|---|---|
| $\rho$ | 0.33 | 0.50 | 0.66 | 0.33 | 0.50 | 0.66 |
| $u_{min}$ | - | - | - | 0.05 | 0.05 | 0.05 |
| $u_{max}$ | - | - | - | 0.15 | 0.15 | 0.15 |
| $u_{min}$ | 0.5 | 0.5 | 0.5 | 0.5 | 0.5 | 0.5 |
| $u_{max}$ | 1.5 | 1.5 | 1.5 | 1.5 | 1.5 | 1.5 |
| $i_{max}$ | 3 | 5 | 7 | 4 | 6 | 10 |
| $MC_{max}$ | 1000 | 1000 | 1000 | 1000 | 5333 | 1706667 |
| $Stdev_{MCmax}$ | 0 | 0 | 0 | 0 | 2309 | 591206 |
| Time (s) | 81 | 127 | 1577 | 114 | 327 | 88307 |
| $Stdev_{Time}$ | 2.1 | 0.58 | 4.4 | 4.0 | 160 | 2290 |

In Table 1, execution parameters for the MATLAB implementation solving the queuing problem for some scenarios are presented. The definitions of scenarios are discussed in the next paragraph. In all cases reported in this table, the method was configured to use 10000 time steps when computing $k_i$ and 5000 time steps for the service time, - to include 99% of the probability mass for the queue length distribution in order to find $i_{max}$, to allow for a 1% tolerance for the integrals of $f_{Wi}(t)$, and to start with the number of Monte Carlo sample points $MC = 1000$.

In the following we present results from two types of scenarios. In the first type, all measurement sessions have the same probe rate distribution, 0.5 – 1.5 Gbps, which means that they all have the same service time distribution. We call a scenario belonging to this type a "pure" scenario.

The second scenario type consists of two groups of measurement sessions, one with a probe rate of 50 Mbps – 150 Mps, and another with 0.5 – 1.5 Gbps. Such a scenario is called a "mixed" scenario.

For both scenarios the packet size $s = 1500$ bytes and the train length $r = 17$.

We have performed MATLAB executions of the method to compute the distributions for several scenarios, with $\rho$ in the range of 0.33 to 0.66 for both the pure and mixed cases. The MATLAB runs have been performed on a standard laptop PC with a Core 2 Duo P8700 processor at 2.53 GHz. As can be seen from Table 1, the number of $f_{Wi}(t)$ terms, the maximum number of Monte Carlo sample points and the execution time increase with $\rho$ and scenario complexity. The computing time increases more and more sharply when $\rho$ approaches 1.

In practice, it is obviously not advisable to stress the queuing system by running it close to the singularity point $\rho = 1$. Still, for the pure scenario, it is feasible to solve the problem and compute the distributions using this setup for rather high $\rho$. The mixed scenario is more challenging, as the structure in the service time distribution spans over a larger time scale, and the same high $\rho$ values cannot easily be explored with the present implementation. From a practical point of view, already the results obtained here indicate that it advisable to avoid mixing measurement sessions with widely varying probe rates in the same queue. The pure scenarios give higher performance both to the computations and to the real system.

First we investigate the queue occupancy. Figure 4a shows the queue occupancy probabilities for $\rho = 0.33$, $\rho = 0.50$ and $\rho = 0.66$, for both pure and mixed scenarios. We can see that the queue length probabilities decrease toward zero for higher values of the queue length, as expected. We can also see that there is a heavier tail for higher values of $\rho$, which is also expected. This means that more queue elements need to be taken into account for higher values of $\rho$ in the computations of the waiting time distribution. This is also shown in Table I, where it can be seen that for the pure scenario $i_{max}$ needs to be 3, 5 and 7, respectively, for $\rho = 0.33, 0.50$ and $0.66$, in order to capture 99% of the probability mass for the queue length distribution. This can also be seen in Figure 4b, showing the cdf of the queue length distribution for these six scenarios.

The queue length distribution can also be seen interleaved in the pdf for the waiting time, as in Figure 5b. However, the dynamics can be seen more clearly when plotting the individual contributions $f_{Wi}(t)\pi_i$ from (20) to the pdf for the waiting time. This is shown in Figure 5a. The figure shows the contributions to the pdf for a pure scenario with $\rho = 0.66$. The area under each component of the pdf corresponds to the probability for a particular queue length. Note that the first component has a smooth curve, since it is plotted as the exact function for the waiting time for one element in the queue, i.e. the service time, which in these scenarios has a $1/t^2$ behavior. The higher components have a shape that reflects the convolution of several elements in the queue plus a fuzzy overlay coming from the statistical noise of the Monte Carlo integrations.

We can also see that the areas of the pdf components in the tail tend to be smaller for smaller $\rho$ values. The pdf components add up to the total pdf for a given value of $\rho$. Figure 5b shows the pdf for the combined impact on the waiting time, for the pure scenario with $\rho = 0.66$. The plotted standard deviation relates to the noise in the pdf coming from the finite sample sizes in the Monte Carlo integrations.

For comparison, Figure 6a shows the mixed scenario with $\rho = 0.66$. In this case the contribution from the two measurement session types is visible. The sessions with a rate between 0.5 and 1.5 Gbps corresponds to the leftmost probability mass, up to around 0.001 seconds, while the rightmost part mainly originates from sessions with rates in [0.05, 0.15] Gbps.

Finally, we compute the cumulative distribution functions for the 6 scenarios, which also enable us to obtain the quantiles of the waiting time distributions. In Figure 6b we see that the cdf converges to probability = 1 for all curves, which of course is a necessary consistency check of the method. For all three pure scenarios it can be seen that e.g. the 90th percentile for the waiting time is below 1 ms.

In the mixed scenarios we see in Figure 6b that there are two main contributors to the cdf, resulting in the "bumps" of the curves. We see that the convergence time is much larger.

Figure 6b also serves as a comparison between the pure and mixed scenario. We can see that the sessions with the lower range of probing rates influence negatively on the percentiles of the waiting times. *This indicates that in a practical measurement system, measurement scenarios where sessions with widely varying probing rates should if possible be handled by different queues, for optimum performance.*

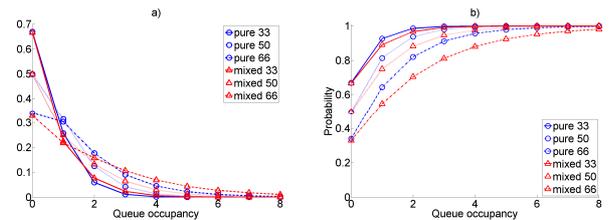

Fig. 4. a) The figure shows the probability mass function for queue occupancy. The distribution is discrete and the sloping lines are plotted to identify the data sets. b) The figure shows the cumulative distribution functions for queue occupancy. The cdf is piece-wise constant; the sloping lines are plotted to identify the data sets.

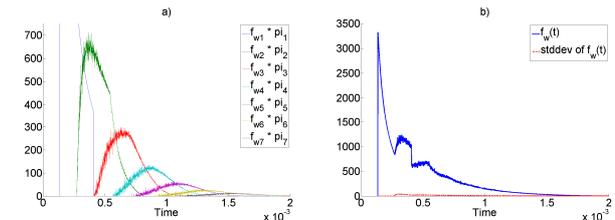

Fig. 5. The left figure (a) is the contributions to the waiting-time pdf from 1, 2, …, 7 queue elements for scenario "Pure 66". The right figure (b) is the total pdf for queue waiting time for scenario "Pure 66".

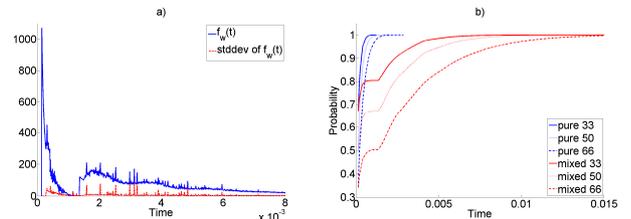

Fig. 6. a) The pdf for queue waiting time for scenario "Mixed 66". b) Cumulative distribution functions for the queue waiting time.

## VIII. Dimensioning of Capacity Measurement System

As seen from the results in the previous section the operational efficiency of the end-point reflector queue is affected by the traffic intensity lambda and the session properties. The numerical results can be used for providing dimensioning guidelines for the waiting time, queue buffer size of a RFC 6802 TWAMP reflector and the maximum allowed intensity, expressed as the total number of received trains per second. One general guideline found in the previous section is that measurement scenarios with sessions having widely varying probing rates should be handled by different queues. Therefore, this section only discusses the dimensioning aspect for "pure" scenarios.

The queue buffer size is an important parameter for the reflector. Further, in order not to overload the reflector the maximum allowed intensity should be a shared parameter among all senders. The waiting time is important from a scheduling perspective. That is, how long round-trip time is expected for a train sent to the reflector?

The framework described in this paper can be used by network operators to determine the queue occupancy and waiting time for a given $\lambda$ which can be seen as the probing budget. To illustrate this we present numerical values based on the $95^{th}$ and $99^{th}$ percentiles for the queue occupancy and waiting time in Table 2 based on different $[u_{max}, u_{min}]$ setting. The parameter $\rho$ is set to approximately 0.9 for the three scenarios. Observe that the parameters $\lambda$ and $\rho$ are directly proportional, as given by (15). Further note that the results are based on "pure" scenarios with only one probe rate interval. Furthermore, all probe traffic in the system has the same size of 1500 bytes and the same train length of 17 packets.

TABLE II. WAITING TIME AND QUEUE LENGTH CALCULATED BASED ON $P = 0.9$ FOR DIFFERENT $[U_{MIN}, U_{MAX}]$ SETTINGS

| $u_{min}$ | $u_{max}$ | $\lambda$ | Waiting time percentile 95 | Waiting time percentile 99 | Queue length percentile 95 | Queue length percentile 99 |
|---|---|---|---|---|---|---|
| 0.5 | 1.5 | 4000 | 3.3 | 4.3 | 16 | 24 |
| 0.05 | 0.15 | 400 | 33.2 | 46.6 | 16 | 24 |
| 0.005 | 0.015 | 40 | 330.5 | 463.2 | 16 | 24 |

For a given scenario and $\lambda$ the operator obtains the queue occupancy and waiting time for the $95^{th}$ and $99^{th}$ percentile.

The method in Section VII is easily deployed when investigating dimensioning aspects where other session parameter settings are used, for example the mixed scenario investigated earlier in the paper.

## IX. Conclusions

This paper outlined an envisioned architecture for large-scale deployment of capacity measurement systems using standard measurement functionality. The paper discusses the scalability related to the end-point nodes of such systems.

The goal was to calculate the queue length and waiting time distributions for the out queue of TWAMP (and similar) reflectors which is the plausible bottleneck in a large-scale system of available path capacity measurements. The results provide a framework for obtaining dimensioning guidelines to compensate for this bottleneck. The guidelines suggest constraints on the total number of concurrent measurement sessions in the system.

An important implication of the results is that in a practical measurement system, measurement scenarios with widely varying probing rates should if possible be handled by different measurement endpoints, for optimum performance.

Note that the framework also can be used for dimensioning the opposite scenario, with a centralized TWAMP Sender, sending probe trains to many different TWAMP reflectors, in case the requests for the probe trains are generated according to the principles discussed in this paper.

To calculate the queue length and waiting time distributions for the out queue of a TWAMP reflector we devised a method, combining an explicit formula for the waiting time probability density function obtained using results from the *M/G/1* queuing theory with the Monte Carlo integration method for computing a multitude of multidimensional convolution integrals. Execution of the method show that the distribution functions of the waiting times can be calculated with reasonable accuracy within reasonable computing times using a modest computational environment.

APPENDIX A - DISTRIBUTION OF SERVICE TIME FOR THE SYSTEM OF N CONCURRENT SESSIONS

We have a queuing system of $N$ concurrent measurement sessions, each with a known intensity $\lambda^{(n)}$ and a known service time distribution ( = train sending time distribution) $f_{T^{(n)}}(t)$.

The total intensity of the system is

$$\lambda = \sum_{n=1}^{N} \lambda^{(n)}$$

Proposition: the distribution of the service time (= train sending time $T$) of the whole system can be obtained as a weighted average of the distributions of the $N$ concurrent measurement sessions:

$$f_T(t) = \frac{1}{\lambda} \sum_{n=1}^{N} \lambda^{(n)} f_{T^{(n)}}(t)$$

Proof: This could be argued from the point of view of conditional probabilities, using the law of total probability:

$$P(A) = \sum_{n=1}^{N} P(A|B^{(n)}) P(B^{(n)})$$

where the $B^{(n)}$ form a partition of the entire sample space (they are mutually exclusive and the union of them is the sample space).

Let $A$ be the event that $t - \varepsilon \leq T \leq t + \varepsilon$, where $\varepsilon > 0$ can be chosen arbitrarily small. We then have

$$P(A) = \int_{t-\varepsilon}^{t+\varepsilon} f_T(u)\, du \approx 2\varepsilon f_T(t)$$

$$f_T(t) \approx \frac{1}{2\varepsilon} P(A)$$

Let $B^{(n)}$ be the event that the train comes from session $n$. We have

$$P(B^{(n)}) = \frac{\lambda^{(n)}}{\lambda}$$

and

$$P(A|B^{(n)}) = \int_{t-\varepsilon}^{t+\varepsilon} f_{T^{(n)}}(u)\, du \approx 2\varepsilon f_{T^{(n)}}(t)$$

Thus,

$$P(A) = \sum_{n=1}^{N} P(A|B^{(n)}) P(B^{(n)}) \approx \sum_{n=1}^{N} 2\varepsilon f_{T^{(n)}}(t) \frac{\lambda^{(n)}}{\lambda}$$

and

$$f_T(t) \approx \frac{1}{2\varepsilon} P(A) \approx \frac{1}{2\varepsilon} \sum_{n=1}^{N} 2\varepsilon f_{T^{(n)}}(t) \frac{\lambda^{(n)}}{\lambda}$$

If we now let $\varepsilon \rightarrow 0$, the errors in the approximations tend to 0, and we then get our result

$$f_T(t) = \frac{1}{\lambda} \sum_{n=1}^{N} \lambda^{(n)} f_{T^{(n)}}(t)$$

APPENDIX B – DERIVATION OF QUEUE LENGTH
DISTRIBUTION FOR A M/G/1 SYSTEM

This derivation is based on [8]. The approach for the queue length distribution uses an embedded discrete-time Markov chain, where the time epochs are immediately after the departures from the queue. The state of the Markov chain at discrete time is the queue occupancy at that time.

From one discrete time to the next, there is by definition exactly 1 departure (unless the queue was empty), and $i$ arrivals ($i = 0, 1, 2, \ldots$). Let $k_i$ be the probability that there are $i$ arrivals during this time.

The transition probability from a given state to the next lower state is then $k_0$, as this transition occurs exactly when there are no arrivals during the service time. Likewise, the transition probability from a state to itself is $k_1$, and the probability for a transition to a state of $j$ units higher occupancy is $k_{j+1}$.

The above holds for all states except for the empty state, where there can be no departure from the queue. In this case, the transition probability from the state to itself is $k_0$, and the probability for a transition to a state of $j$ units higher occupancy is $k_j$.

For all transitions from one state to a state of 2 or more units lower the probability is always zero, since the discrete time is defined by only 1 departure.

With the transition probabilities from state $l$ to state $m$ in row $l$ and column $m$, ($l = 0, 1, 2,\ldots; m = 0, 1, 2, \ldots$) we can write the state transition matrix:

$$P = \begin{pmatrix} k_0 & k_1 & k_2 & k_3 & \cdots \\ k_0 & k_1 & k_2 & k_3 & \cdots \\ 0 & k_0 & k_1 & k_2 & \cdots \\ 0 & 0 & k_0 & k_1 & \cdots \\ \vdots & \vdots & \vdots & \vdots & \ddots \end{pmatrix}$$

From the standard theory of Markov chains, we can find the steady-state probabilities

$$\pi = [\pi_0 \quad \pi_1 \quad \pi_2 \quad \pi_3 \ldots]$$

where $\pi_i$ is the probability that the system is in state $i$, i.e. that the queue length is $i$, from the equation $\pi = \pi P$. Each component can then be calculated as

$$\pi_i = \pi_0 k_i + \sum_{j=1}^{i+1} \pi_j k_{i-j+1} \quad (i = 0, 1, 2, \ldots)$$

We can rearrange the above equation as

$$\pi_{i+1} k_0 = \pi_i - \pi_0 k_i - \sum_{j=1}^{i} \pi_j k_{i-j+1} \quad (i = 0, 1, 2, \ldots)$$

which constitutes a recursion equation for the steady-state queue length probabilities $\pi_1, \pi_2, \pi_3, \ldots$

This formulation allows us to iteratively compute the occupancy probabilities. The starting point for the computation is the well-known probability of an empty queue $\pi_0 = 1 - \rho$.

The probability $k_i$ that there are $i$ arrivals during a service time $T$ can be computed from the number of Poisson arrivals conditioning on $T = t$

$$k_i = \int_0^\infty \frac{e^{-\lambda t}(\lambda t)^i}{i!} f_T(t)\, dt$$

This means that we can now compute as many $\pi_i$ as we need. The cdf $\Pi_i$ for the queue length distribution can now easily be computed from $\pi_i$ as a cumulative sum.

$$\Pi_i = \sum_{j=0}^{i} \pi_j$$

It should be noted that even if we have calculated the steady-state queue length probabilities at the departure epochs, the result is in fact valid for an arbitrary time, due to the PASTA property that Poisson arrivals see time averages. This is for example shown in [8].